\documentclass{article}\usepackage{jkas2}
\usepackage{txfonts}
\usepackage{graphicx}
\usepackage{natbib}

\runningauthor{Pfrommer \& En{\ss}lin}
\runningtitle{The quest for cosmic ray protons in galaxy clusters}
\beginpage{1}
\endpage{5}

\begin{document}

\font\twelvei = cmmi10 scaled\magstep1 
       \font\teni = cmmi10 \font\seveni = cmmi7
\font\mbf = cmmib10 scaled\magstep1
       \font\mbfs = cmmib10 \font\mbfss = cmmib10 scaled 833
\font\msybf = cmbsy10 scaled\magstep1
       \font\msybfs = cmbsy10 \font\msybfss = cmbsy10 scaled 833
\textfont1 = \twelvei
       \scriptfont1 = \twelvei \scriptscriptfont1 = \teni
       \def\mit{\fam1 }
\textfont9 = \mbf
       \scriptfont9 = \mbfs \scriptscriptfont9 = \mbfss
       \def\bmit{\fam9 }
\textfont10 = \msybf
       \scriptfont10 = \msybfs \scriptscriptfont10 = \msybfss
       \def\bmsy{\fam10 }

\def\lsim{\raise0.3ex\hbox{$<$}\kern-0.75em{\lower0.65ex\hbox{$\sim$}}}
\def\gsim{\raise0.3ex\hbox{$>$}\kern-0.75em{\lower0.65ex\hbox{$\sim$}}}

\newcommand{\degr}{\hbox{$^\circ$}}
\newcommand{\crp}{\mathrm{CRp}}
\newcommand{\rmn}{\mathrm}

\title{The quest for cosmic ray protons in galaxy clusters}

\author{C. Pfrommer and T. A. En{\ss}lin}
\address{Max-Planck-Institut f\"{u}r Astrophysik,
Karl-Schwarzschild-Str.1, PO Box 1317, 85741 Garching, Germany \\
{\it E-mail: pfrommer@mpa-garching.mpg.de, ensslin@mpa-garching.mpg.de}}

%\author{and}
%
%\author{...}
%\address{...}

%\author{...$^{1}$ and ...$^{2}$}
%\address{$^{1}$ ...}
%\address{$^{2}$ ...}

%\author{~}
\address{\normalsize{\it (Received October 31, 2004; Accepted December 1, 2004)}}

\abstract{There have been many speculations about the presence of cosmic ray
  protons (CRps) in galaxy clusters over the past two decades. However, no
  direct evidence such as the characteristic $\gamma$-ray signature of decaying
  pions has been found so far. These pions would be a direct tracer of hadronic
  CRp interactions with the ambient thermal gas also yielding observable
  synchrotron and inverse Compton emission by additionally produced secondary
  electrons. The obvious question concerns the type of galaxy clusters most
  likely to yield a signal: Particularly suited sites should be cluster cooling
  cores due to their high gas and magnetic energy densities.  We studied a
  nearby sample of clusters evincing cooling cores in order to place stringent
  limits on the cluster CRp population by using non-detections of {\em EGRET}.
  In this context, we examined the possibility of a hadronic origin of
  Coma-sized radio halos as well as radio mini-halos.  Especially for
  mini-halos, strong clues are provided by the very plausible small amount of
  required CRp energy density and a matching radio profile.  Introducing the
  {\em hadronic minimum energy criterion}, we show that the energetically
  favored CRp energy density is constrained to $2\% \pm 1\%$ of the thermal
  energy density in Perseus.  We also studied the CRp population within the
  cooling core region of Virgo using the TeV $\gamma$-ray detection of M~87 by
  {\em HEGRA}.  Both the expected radial $\gamma$-ray profile and the required
  amount of CRp support this hadronic scenario.}


\maketitle

%\section {Introduction}
\section{Galaxy clusters as a window to non-thermal intra-cluster medium
  components}
\label{Pfrommer:intro}

Cooling cores in galaxy clusters are especially well suited places to find
traces of otherwise nearly invisible non-thermal components of the
intra-cluster medium (ICM) due to the extreme gas densities observed in these
central regions. The faded and therefore invisible remnants of radio galaxy
cocoons, so-called {\it radio ghosts} (En{\ss}lin 1999) or {\it ghost
  cavities}, were first detected in cooling cores by the absence of X-ray
emissivity in the ghost's volume in contrast to the highly X-ray luminous
cooling cores gas surrounding it (B{\"o}hringer et al. 1993; Fabian et al.
2000, and many recent Chandra observations).  Cosmic ray electrons (CRes) are
seen in cooling cores by their radio synchrotron radiation within strong
magnetic fields. Cosmic ray protons (CRps) in the ICM are most likely to be
detected for the first time within cooling cores via their hadronic interaction
with the dense cooling core gas leading to $\gamma$-rays and CRes.

A better knowledge of these non-thermal components of the ICM -- especially in
the cooling core regions -- is highly desirable, since they play important
roles in the heat balance of the gas through heating by CRps, radio ghost
buoyant movements, and suppression of heat conduction by magnetic fields.
Additionally, such non-thermal components are tracers of the violent dynamics
of the ICM and may help to solve some of the puzzles about cooling cores.

In this article, we present our recent progress in constraining CRps by using
limits on the $\gamma$-ray emission as well as the observed radio emission of
nearby galaxy clusters (Sect.~\ref{Pfrommer:gamma} and
Sect.~\ref{Pfrommer:mini-halos}). Finally, we present the novel concept of the
hadronic minimum energy condition which should serve as unambiguous energetic
criterion for the applicability of the hadronic model of radio synchrotron
emission in galaxy clusters (Sect.~\ref{Pfrommer:MEC}).  We focus on ideas and
results, leaving the technical details to the publications given in the
reference list.

\section{Hadronic interactions of cosmic ray protons}
\label{Pfrommer:CRp}
Approximately once in a Hubble time, a CRp collides inelastically with a nucleon
of the ambient ICM gas of non-cooling core clusters. Within cooling cores, such
collisions are much more frequent due to the higher target densities. Such
inelastic proton ($p$) nucleon ($N$) collisions hadronically produce secondary
particles like relativistic electrons, positrons, neutrinos and $\gamma$-rays
according to the following reaction chain:
\begin{eqnarray}
p + N &\rightarrow& 2N + \pi^{\pm/0} 
\nonumber\\
  \pi^\pm &\rightarrow& \mu^\pm + \nu_{\mu}/\bar{\nu}_{\mu} \rightarrow
  e^\pm + \nu_{e}/\bar{\nu}_{e} + \nu_{\mu} + \bar{\nu}_{\mu}\nonumber\\
  \pi^0 &\rightarrow& 2 \gamma \,.\nonumber
\end{eqnarray}
The resulting $\gamma$-rays can be detected directly with current and future
$\gamma$-ray telescopes. The relativistic electrons and positrons (summarized as
CRes) are visible due to two radiation processes: inverse Compton scattering of
background photon fields (mainly the cosmic microwave background, but also
starlight photons) and radio synchrotron emission in ICM magnetic fields.
Especially the latter process provides a very sensitive observational signature
of the presence of CRps in cooling cores not only because of the tremendous
collecting area of radio telescopes, but also due to the strong magnetic fields
in cooling cores, as Faraday rotation measurements demonstrate.

\section{Gamma-ray emission}
\label{Pfrommer:gamma}

\subsection{Gamma-ray constraints from {\em EGRET}}
\label{Pfrommer:EGRET}

\begin{figure}[t]
\resizebox{\hsize}{!}{
\includegraphics{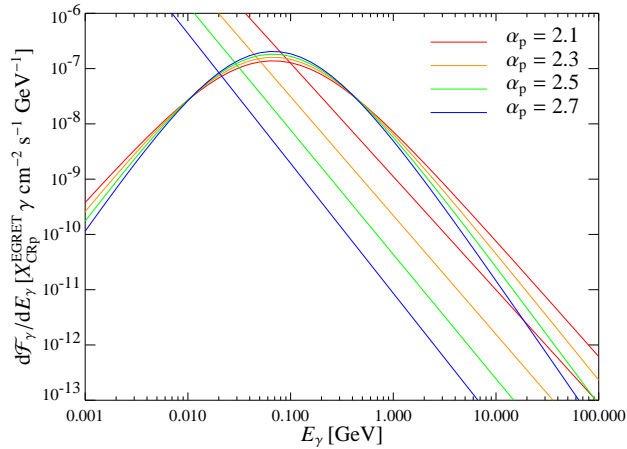}}
\caption{Theoretical $\gamma$-ray spectra of the Perseus cluster resulting from
  CRp hadronic interactions with the cooling core gas.  The assumed CRp spectra
  are normalized to be marginally consistent with the {\em EGRET} $E_\gamma >
  100$ MeV non-detection limits. Results for different proton spectral indices
  are shown. The $\pi^0$-decay bump is clearly visible. Secondary CRes scatter
  cosmic microwave background photons into the $\gamma$-ray regime by the
  inverse Compton process. The resulting power-law emission dominates the low
  energy part of the spectrum.  The CRe population was calculated neglecting
  synchrotron radiation energy losses. Thus, realistic inverse Compton spectra
  due to this process will exhibit a lower normalization than displayed here.}
\vspace{-0.5cm}
\label{Pfrommer:Perseus}
\end{figure}

\begin{figure}[t]
\resizebox{\hsize}{!}{
\includegraphics{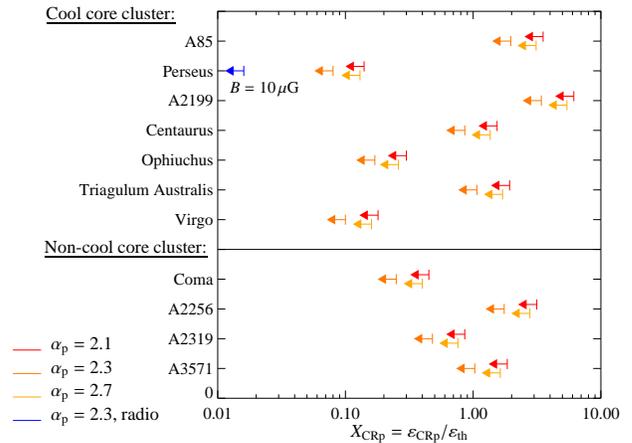}}
  \caption{{\em EGRET} $\gamma$-ray constraints on the (central) CRp energy density
    ($\varepsilon_{\rm CRp}$, for various CRp spectral indices) in terms of the
    cluster central thermal energy density ($\varepsilon_{\rm th}$) for cooling
    core and non-cooling core clusters.  The constraint from the Perseus
    cluster radio mini-halo observation is also displayed while assuming a
    typical cooling core cluster magnetic field strength (blue arrow). If radio
    mini-halos are entirely produced hadronically by CRps then the constraint
    derived for the Perseus cluster is an actual measurement for
    $X_\mathrm{CRp}$ and not only an upper limit.}
\label{Pfrommer:gammaConstr}
\end{figure}

Assuming that the CRp population can be described by a power law distribution
in momentum, Pfrommer \& En{\ss}lin (2004a) developed an analytical formalism to
describe the secondary emission spectra from hadronic CRp interactions which
exhibit the simplicity of textbook formulae. This formalism was applied to a
sample of nearby X-ray luminous galaxy clusters which are also believed to be
powerful $\gamma$-ray emitters owing to the present high target densities.
Synthetic $\gamma$-ray spectra of the Perseus cooling core cluster, which are
calculated using this formalism, are shown in Fig.~\ref{Pfrommer:Perseus}.
Assuming that the CRp population follows the spatial distribution of the
thermal ambient intra-cluster gas, we can define the CRp scaling ratio
\begin{equation}
  \label{eq:X_CRp}
  X_\mathrm{CRp} \equiv
  \frac{\varepsilon_\mathrm{CRp}}{\varepsilon_\mathrm{th}}, 
\end{equation}
where $\varepsilon_\mathrm{CRp}$ and $\varepsilon_\mathrm{th}$ denote the CRp
and thermal energy density, respectively. This scaling effectively implies an
{\em isobaric} compression of the CRps during the formation of the cooling
core.  The parent CRp spectra, which give rise to the hadronically induced
$\gamma$-ray spectra, are normalized to upper limits on the $\gamma$-ray
emission obtained by {\em EGRET} observations for energies $E_\gamma > 100$ MeV
(Reimer et al. 2003). For nearby cooling core clusters, this analysis
constrains CRp energy densities relative to the thermal energy density to
$X_\mathrm{CRp} \sim 10\%$.  For the full sample of nearby X-ray luminous
clusters, upper limits of the same order of magnitude were obtained (cf.{\ 
  }Fig.~\ref{Pfrommer:gammaConstr}). It is obvious from this compilation that
cooling core clusters are extremely well suited to visualize even small CRp
populations.

The real $\gamma$-ray spectra are not expected to be far below these
spectra: many processes like supernova driven galactic winds, structure
formation shock waves, radio galaxy activity, and in-situ turbulent particle
acceleration support this expectation. These processes should have produced a
CRp population characterized by an energy density relative to the thermal
energy density of at least a few percent.

\subsection{Gamma-rays from the central region of the Virgo cluster}
\label{Pfrommer:virgo}

\begin{figure}[!t]
\resizebox{\hsize}{!}{
\includegraphics{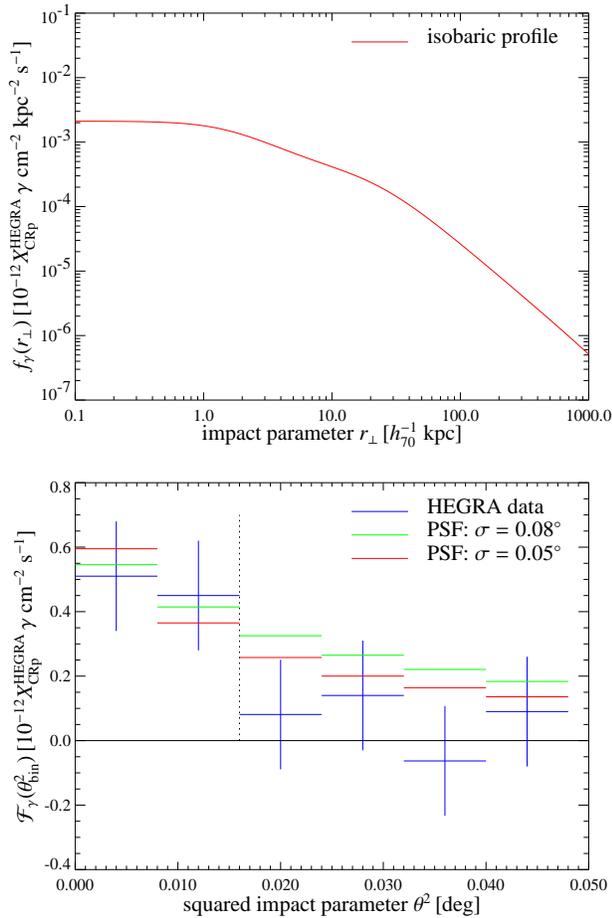}}
  \caption{{\bf Upper panel:} Modeled $\gamma$-ray surface flux profile of the
    cooling core region within the Virgo cluster. The profile is normalized by comparing
    the integrated $\gamma$-ray flux above 730~GeV to {\em HEGRA} data of M~87 within
    the innermost two data points.  {\bf Lower panel:} Comparison of detected
    to predicted $\gamma$-ray flux within the central aperture and the
    innermost annuli for two different widths of the PSF. The red lines
    correspond to an optimistic PSF of $\sigma = 0.05\degr$ whereas the green
    lines are calculated for a conservative $\sigma = 0.08\degr$ for a soft
    $\gamma$-ray spectrum. The vertical dashed line separates the data from the
    noise level at a position corresponding to $r_\bot = 37.5$~kpc.}
\label{Pfrommer:Virgo}
\end{figure}

Recently, the {\em HEGRA} collaboration (Aharonian et al. 2003) announced
a TeV $\gamma$-ray detection from the giant elliptical galaxy M~87 which
is situated at the center of the cooling core region of the Virgo cluster.
Using imaging atmospheric \v{C}erenkov techniques, this $\gamma$-ray
detection was obtained at a 4-$\sigma$ significance level. On the
basis of their limited event statistics, it is inconclusive whether
the detected emission originates from a point source or an extended
source. Despite testing for intermittent behavior of M~87, no time
variation of the TeV $\gamma$-ray flux has been found.

Pfrommer \& En{\ss}lin (2003) applied the previously described
analytical formalism (Sect.~\ref{Pfrommer:EGRET}) of secondary $\gamma$
ray emission spectra resulting from hadronic CRp interactions to the
central cooling core region of the Virgo cluster. While combining the observed
TeV $\gamma$-ray emission to {\em EGRET} upper limits, they constrain the CRp
spectral index $\alpha_\mathrm{GeV}^\mathrm{TeV}<2.3$ provided the
$\gamma$-ray emission is of hadronic origin and the population is
described by a single power-law ranging from the GeV to TeV energy
regime. 

A comparison of the observed to the predicted $\gamma$-ray emissivity
profiles is shown in Fig.~\ref{Pfrommer:Virgo}. In order to allow for
finite resolution of the \v{C}erenkov telescope, the real profile was
convolved with the point spread function (PSF) of the {\em HEGRA}
instrument. The more optimistic PSF of $\sigma = 0.05\degr$ for harder
$\gamma$-ray spectra being favored by our model as well as the
conservative choice of a PSF derived from softer Crab-like spectra are
both consistent with the observed data. Considering an aged CRp
population which has already suffered from significant Coulomb losses,
CRp scaling ratios of the order of $X_\mathrm{CRp} \sim 50\%$ are
obtained for the innermost region of Virgo within $r_\bot = 37.5$~kpc.

Since the emission region is dominated by the giant radio galaxy M~87, other
mechanisms like processed radiation of the relativistic outflow or dark matter
annihilation could also give rise to the observed $\gamma$-ray emission.
Nevertheless, the hadronic scenario probes the CRp population within a mixture
of the inter stellar medium of M~87 and the ICM of the center of Virgo yielding
either an upper limit or a detection on the CRp population, provided this
scenario applies.

\section{Radio mini-halos}
\label{Pfrommer:mini-halos}

\begin{figure}[t]
\resizebox{\hsize}{!}{
\includegraphics{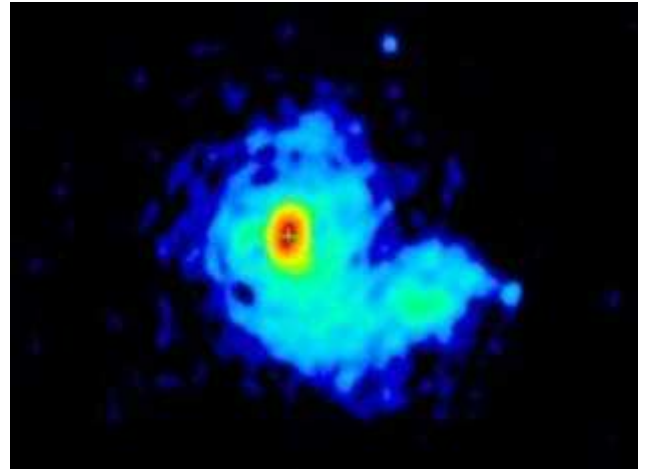}}
  \caption{The Perseus radio mini-halo at 1.4 GHz from
    Pedlar et al.~(1990) with an extent of $~160 \,h_{70}^{-1}$~kpc in
    diameter. At the very center is an extremely bright flat-spectrum core
    owing to relativistic outflows of the radio galaxy NGC 1275. This outflow
    was subtracted from this image to make the extended structure more visible.
    The position of the radio galaxy is marked by the green cross in the center
    of the red region.}
\label{Pfrommer:radioPerseuspic}
\end{figure}

\begin{figure}[t]
  \resizebox{\hsize}{!}{
\includegraphics{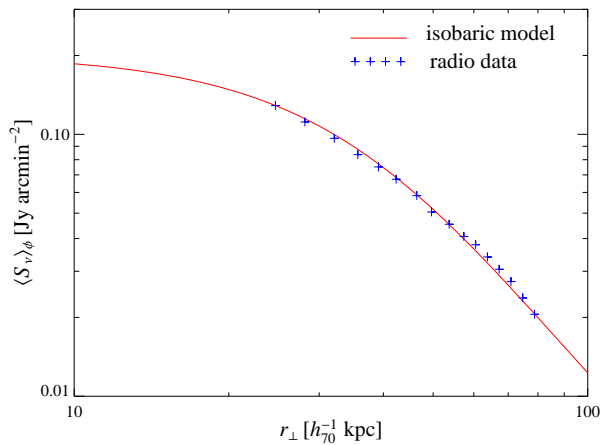}}
  \caption{Radio emissivity profile of the Perseus mini-halo at 1.4 GHz
    observed by Pedlar et al.~(1990) (crosses) and predicted due to
    hadronic interactions of CRp by Pfrommer \& En{\ss}lin (2004a). Both the
    required CRp scaling ratio $X_\mathrm{CRp} \simeq 2\%$ and the
    morphological match of the observed and predicted radio emissivities
    strongly indicate the hadronic origin of radio mini-halos.}
\label{Pfrommer:radioPerseus}
\end{figure}

The Perseus radio mini-halo at 1.4~GHz is shown in
Fig.~\ref{Pfrommer:radioPerseuspic} as observed by Pedlar et al.~(1990).
The emission due to the relativistic jet has been subtracted to make the
extended structure more visible. The spatial extent of the radio mini-halo of
$~160 \,h_{70}^{-1}$~kpc in diameter is too large to be accounted for by
synchrotron emission of direct accelerated CRes in structure formation or
accretion shocks. Thus one needs to consider other injection processes of CRes
into the ICM which are responsible for the observed diffuse radio emission.
Besides the reacceleration scenario of mildly relativistic CRes ($\gamma\simeq
100-300$) which are accelerated in-situ by turbulent Alv\'en waves
(Gitti et al. 2002), the hadronic injection scenario of CRes is a
promising alternative.

Azimuthally averaging the radio emission shown in
Fig.~\ref{Pfrommer:radioPerseuspic} yields the radio emissivity profile as
displayed in Fig.~\ref{Pfrommer:radioPerseus}.  For comparison, we overlaid
synthetic radio surface brightness profiles resulting from the hadronic
scenario. There is a perfect morphological match between observed and predicted
profiles.  Assuming a typical magnetic field strength for cooling cores of
$10\,\mu$G, we derive an upper limit for the CRp population of $X_\mathrm{CRp}
\simeq 2\%$ by normalizing our simulated to the observed profile (cf.{\ 
  }Fig.~\ref{Pfrommer:gammaConstr}).

The radio emissivity of hadronically generated CRes depends on the assumed
magnetic field strength. Thus, upper limits on $X_\mathrm{CRp}$ rely on the
same assumption. However, in the case of strong magnetic fields (above
$3\,\mu$G), the dependence of $X_\mathrm{CRp}$ on the assumed field strength is
very weak, since synchrotron losses dominate in this regime. If radio
mini-halos are entirely produced hadronically by CRps, then the constraint
derived for the Perseus cluster is an actual measurement for $X_\mathrm{CRp}$
and not only an upper limit.  Alongside the perfect morphological match of the
radio synchrotron profiles, the comparably small number of CRps required to
account for the observed radio mini-halo suggests a hadronic origin of the
radio mini-halo in Perseus.

\section{Hadronic minimum energy criteria}
\label{Pfrommer:MEC}

\begin{figure*}[t]
\begin{tabular}{cc}
\resizebox{0.48\hsize}{!}{\includegraphics{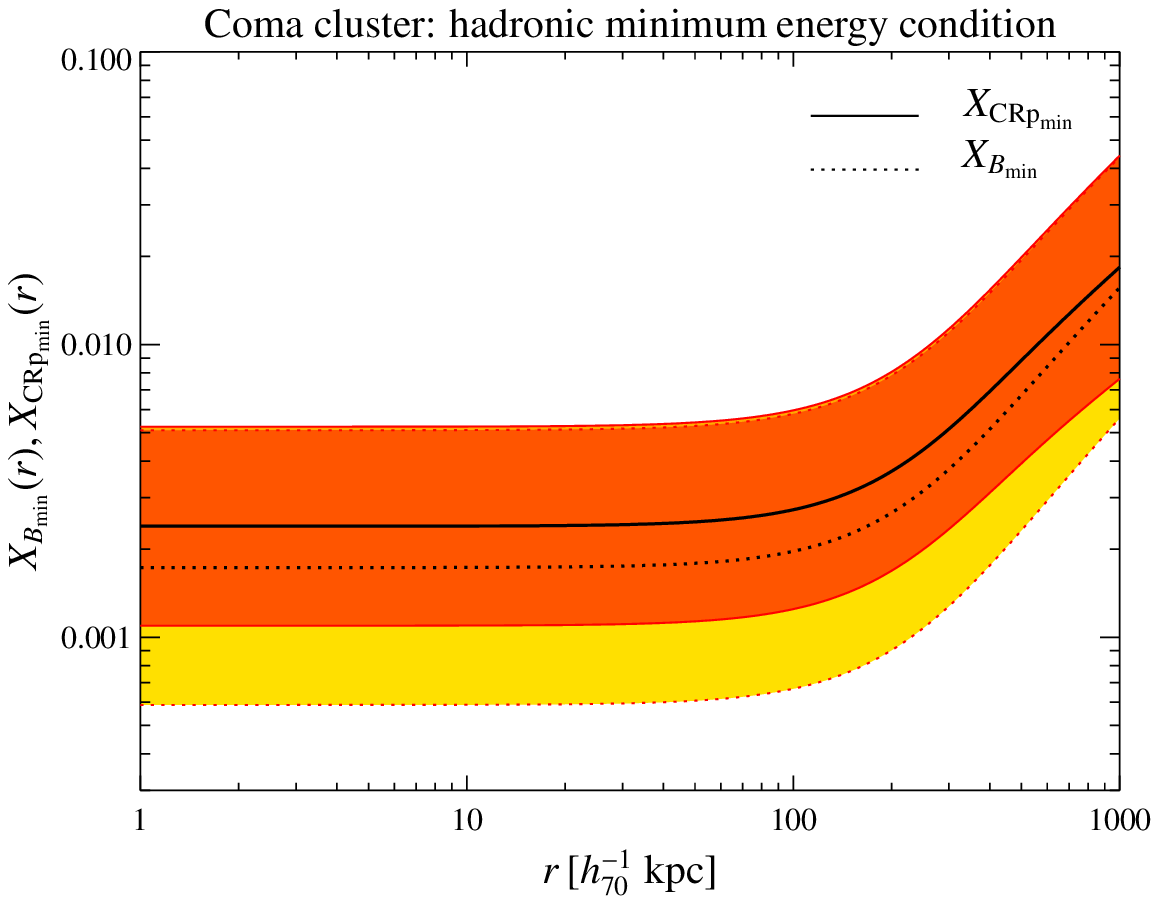}} &
\resizebox{0.48\hsize}{!}{\includegraphics{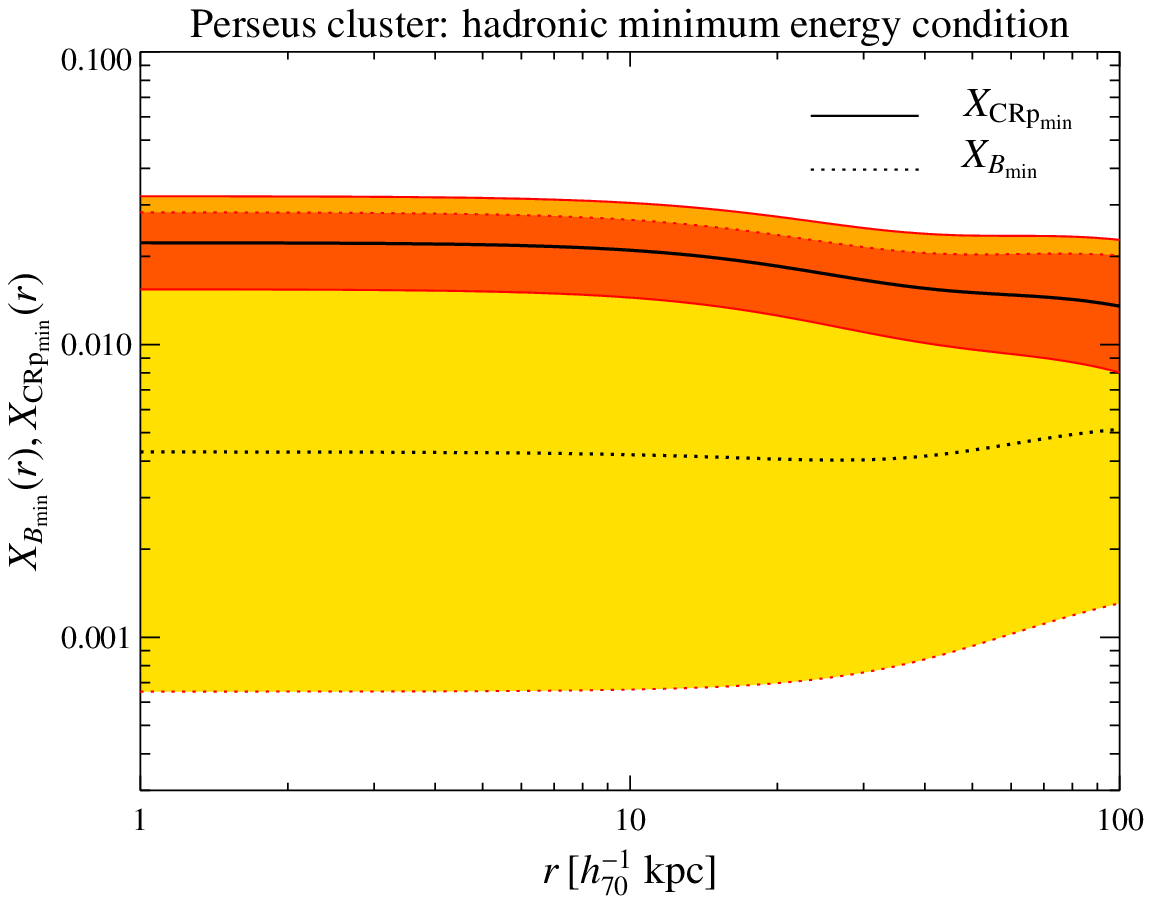}}
\end{tabular}
  \caption{Profiles of the CRp-to-thermal energy density
    $X_{\crp_\rmn{min}}(r)$ (solid) and magnetic-to-thermal energy density
    $X_{B_\rmn{min}}(r)$ (dotted) as a function of deprojected radius are
    shown.  The different energy densities are obtained by means of the
    hadronic minimum energy criterion.  The {\bf left panel} shows profiles of
    the Coma cluster while the {\bf right panel} represents profiles of the
    Perseus cluster.  The light shaded areas represent the logarithmic
    tolerance regions of $X_{B_\rmn{min}}(r)$ and $X_{\crp_\rmn{min}}(r)$,
    respectively, while the dark shaded regions indicate the overlap and thus
    the possible equipartition regions in the quasi-optimal case.}
\label{Pfrommer:MECfig}
\end{figure*}

As an extension of the previous work, we estimated magnetic field strengths of
radio emitting galaxy clusters by minimizing the non-thermal energy density ---
contained in relativistic electrons, protons, and magnetic fields --- with
respect to the magnetic field strength (Pfrommer \& En{\ss}lin 2004b).  As one
boundary condition, the implied synchrotron emissivity is required to match the
observed value. Additionally, a second boundary condition is required
mathematically which couples CRps and CRes.  For the {\em classical} case, a
constant scaling factor between CRp and CRe energy densities is assumed.
However, if the physical connection between CRps and CRes is known or assumed,
a physically better motivated criterion can be formulated.  As such a case, we
introduce the minimum energy criterion within the scenario of {\em
  hadronically} generated CRes. The work by Pfrommer \& En{\ss}lin (2004b)
provides simple self-consistent recipes for applying the classical and hadronic
minimum energy criterion in typical observational situations.

Alongside, we provide theoretically expected tolerance regions which measure
the deviation from the minimum energy states by one e-fold: We use logarithmic
measures of the curvature radius at the extremal values in order to
characterize the `sharpness' of the minima.  These regions have the meaning of
a quasi-optimal realization of the particular energy densities.

The philosophy of this approach is to provide a criterion for the energetically
least expensive radio synchrotron emission model possible for a given
physically motivated scenario.  There is no first principle enforcing this
minimum state to be realized in nature. However, our minimum energy estimates
are interesting in two respects: First, these estimates allow scrutinizing the
hadronic model for extended radio synchrotron emission in clusters of galaxies.
If it turns out that the required minimum non-thermal energy densities are too
large compared to the thermal energy density, the hadronic scenario will become
implausible to account for the extended diffuse radio emission. In this
respect, our criteria is a way to test the robustness of the previous results
with respect to the observationally available parameter space spanned by the
CRp spectral index and the (unknown) distribution of the magnetic field
strength.  Secondly, should the hadronic scenario be confirmed, the minimum
energy estimates allow testing for the realization of the minimum energy state
for a given independent measurement of the magnetic field strength.

Application to the radio halo of the Coma cluster (Deiss et al. 1997) and the
radio mini-halo of the Perseus cluster (Pedlar et al. 1990) yields
equipartition between cosmic rays and magnetic fields within the expected
tolerance regions. In the hadronic scenario, the inferred central magnetic
field strength ranges from $2.4~\mu\rmn{G}$ (Coma) to $8.8~\mu\rmn{G}$
(Perseus), while the optimal CRp energy density is constrained to $2\% \pm 1\%$
of the thermal energy density (Perseus) (cf.~Fig.~\ref{Pfrommer:MECfig}).
Pfrommer \& En{\ss}lin (2004b) discuss the possibility of a hadronic origin of
the Coma radio halo while current observations favor such a scenario for the
Perseus radio mini-halo.  Combining future expected detections of radio
synchrotron, hard X-ray inverse Compton, and hadronically induced $\gamma$-ray
emission should allow an estimate of volume averaged cluster magnetic fields
and provide information about their dynamical state.

\section{Conclusions}

\begin{itemize}
 \item We argue that cooling cores of galaxy clusters are well suited
 to reveal or constrain any cosmic ray proton population via radiation
 from hadronic interactions with the ambient gas nuclei. Such
 collisions lead to $\gamma$-rays and cosmic ray electrons. The former
 would have been seen above 100 MeV by the {\em EGRET} telescope if the
 cosmic ray protons had energy densities relative to the thermal gas
 exceeding 10\% (Pfrommer \& En{\ss}lin 2004a).
\item The giant elliptical galaxy M~87 in the center of the Virgo cluster
  cooling core region has recently been detected at TeV energies by the {\em
    HEGRA} instrument (Aharonian et al. 2003). These $\gamma$-rays could be
  produced by hadronic interactions of a cosmic ray proton population if its
  spectral index $\alpha_\mathrm{GeV}^\mathrm{TeV}<2.3$ and its energy density
  is of the order of $50\%$ of the gas within the transition/mixture of
  inter-stellar and intra-cluster medium within M~87 (Pfrommer \& En{\ss}lin
  2003).
 \item Cosmic ray electrons produced in hadronic interactions are a
very sensitive indicator of cosmic ray protons due to their strong
emissivity. Radio synchrotron emission of such electrons in strong
cooling core magnetic fields of $\sim 10\,\mu$G limit the cosmic ray
proton energy density to $\sim 2\%$ or less compared to the thermal one in
the Perseus cluster cooling core (Pfrommer \& En{\ss}lin 2004a).
\item Diffuse radio emission from the Perseus cluster cooling core was detected
  -- the so called Perseus {\it radio mini-halo}. This radio synchrotron
  emission may be induced by cosmic ray proton interactions in the
  intra-cluster medium. This
  scenario is strongly supported by the very moderate energy requirements (2\%
  of the thermal energy) and the excellent agreement between the observed and
  the theoretically predicted radio surface profile (Pfrommer \& En{\ss}lin 2004a).
\item Introducing the {\em hadronic minimum energy criterion}, we show that the
  energetically favored cosmic ray proton energy density is constrained to $2\%
  \pm 1\%$ of the thermal energy density in Perseus. Application to the radio
  halo of the Coma cluster and the radio mini-halo of the Perseus cluster
  yields equipartition between cosmic rays and magnetic fields within the
  expected tolerance regions (Pfrommer \& En{\ss}lin 2004b).
\end{itemize}

\acknowledgements{We thank Francesco Miniati, Bj{\"o}rn Malte Sch{\"a}fer, and
  Sebastian Heinz for discussion and collaboration, Alan Pedlar for the
  permission to reproduce the Perseus mini-halo image, and the conference
  organizers for an excellent meeting.}

\end{document}